\documentclass[3p, twocolumn,times,authoryear]{elsarticle}

\usepackage{hyperref}

\journal{Icarus}

\biboptions{authoryear}

\begin{document}

\begin{frontmatter}

\title{Constraints on Compound Chondrule Formation from Laboratory High-Temperature Collisions}

\author{Tabea Bogdan}
\author{Jens Teiser}
\author{Nikolai Fischer}
\author{Maximilian Kruss}
\author{Gerhard Wurm}
\ead{gerhard.wurm@uni-due.de}

\address{University of Duisburg-Essen, Faculty of Physics, Lotharstr. 1-21, 47057 Duisburg, Germany}

\begin{abstract}
In laboratory experiments, spherical 1\,mm-wide glass and basalt particles are heated, and the hot particles collide at about 1\,m/s with a flat glass target that is at room temperature. When the particles are heated below 900\,K, the collisions are essentially elastic with coefficients of restitution of about 0.9, but above 900\,K collisions become increasingly inelastic and the coefficient of restitution decreases with increasing temperature. At 1100\,K the glass particles approach sticking but, simultaneously, at the same temperature the particles melt on timescales of minutes. The basalt particles approach sticking at 1200\,K. Only above 1400\,K do basalt grains in contact with each other fuse together, forming compounds on timescales of hours, and at 1500\,K basalt grains completely fuse together. Therefore, cooling basalt grains only have a 100\,K window for compound formation, and velocities very likely have to be below 1\,m/s for sticking in the first place. We predict that this puts constraints on compound chondrule formation and particle densities in the solar nebula.
\end{abstract}


\end{frontmatter}


\section{Introduction}

Chondrules are a major component in primitive meteorites. They are millimeter-sized, spherical grains that were flash molten in early phases of planet formation \citep{Alexander2001, Hewins2011}.
Most of them are thought to be meltings of fine-grained dust present in the solar nebula, although interaction with the nebular gas might also have played a role during their formation \citep{Libourel2006}.
Mechanisms for their formation suggested in the past include shocks \citep{Hood1991, Iida2001, Desch2002, Ciesla2002}, condensation in a vapor plume of colliding planetary embryos \citep{Krot2007}, planetesimal collisions \citep{Asphaug2011, Johnson2015, Lichtenberg2018}, magnetic reconnection \citep{Ebel2012,McNally2014}, lightning \citep{Desch2000} or X-winds \citep{Liffman1996}. 
The typical peak temperatures found in the literature are 1500 to 1600\,$^{\circ}$C or 1800 to 1900\,K \citep{Radomsky1990, Zanda2004}. All heating events mentioned are rather short, and chondrules rapidly cool afterwards, and the deduced cooling rates are between 10 and 1000\,K/h \citep{Radomsky1990, Zanda2004}. 

Particles collide all the time in the solar nebula, and valuable evidence of this comes from the collision of hot chondrules. If particles collide (partially) molten they can form compound chondrules that are more or less fused together. Hot collisions are at least one of the widely spread models on compound chondrule formation. \citet{Ciesla2004}, e.g. use the fraction of compounds found in meteorites to constrain the particle number density of chondrules in the solar nebula. The idea is that only a limited time is available for fusing collisions while the molten chondrules cool. Therefore, if the number density is low, collisions are rare during this time, and only a small amount of compounds should be found. For higher-number densities, more compounds should be present. An extreme case of a compound chondrule with 16 individual chondrules stuck together has recently been studied by \citet{Bischoff2017}. This compound was a combination of siblings and independents. Siblings have clearly been generated in the same heating event, and their texture implies that they collided while they were still hot. For independents, it might have been sufficient that one chondrule was hot enough to fuse with the other one. \mbox{\citet{Arakawa2016}} proposed that compounds might also form from supercooled melts, which would extend the possible range of temperatures. A somewhat extreme assumption in terms of temperatures needed to form a compound is discussed by \citet{Hubbard2015}. He argues that chondrules might stick together when cold and only fuse with each other later while being heated only moderately to not much more than 1025 K.  \citet{Miura2008} and  \citet{Yasuda2009} specifically assume shocks as a melting event for a centimeter-sized aggregate and a subsequent stripping of the molten outer layer into fragments that collide again afterwards, forming compounds. Other ideas on compound chondrule formation are mentioned below. In this paper, however, we focus on the idea of fusing chondrules together in collisions of hot grains.

To determine the probability of compound chondrule formation in detail, one would need to consider (1) a specific cooling curve, (2) the exact temperatures that lead to sticking at different collision velocities, (3) the evolution of collision velocities after a chondrule-forming event, i.e., if collisions lead to sticking and (4) the temperature range allowing chondrules in contact with each other to fuse together during cooling. All of these aspects are not well constrained yet, however, and laboratory experiments on collisional formation are rare \citep{Connolly1994}. At the fluid end, \citet{Ashgriz1990} studied coalescence and fragmentation in collisions of water droplets. \citet{Qian1997} have also studied water droplets and hydrocarbon fluids, varying the ambient gas pressure, gas species and impact parameters. Results from both studies were interpreted in terms of the Weber number. Here, we also approach compound formation in laboratory collision experiments but with the focus on collisions  of glass and basalt particles at high temperatures at the transition between solid and liquid.

Here, we define the formation of a compound chondrule as occurring when two spheres fuse together and form significant necks with about 1/3 of the particle diameter (see below for details).

In protoplanetary disks, millimeter-sized particles usually collide at velocities below 1\,m/s \citep{Blum2008}. Their collision velocity might be increased one way or the other through the shocks, gas expansion or photophoresis, especially after a melting event \citep{Wurm2009,Wurm2010,Loesche2016}.
It has generally been taken for granted that chondrules that collide at hot temperatures stick together even at collision velocities of 1\,m/s \citep{Kring1991, Ciesla2004}. \citet{Bischoff2017} even assume that higher collision velocities of 5\,m/s lead to sticking. So far, however, sticking velocities and collisional energy dissipation for high temperatures are only achieved through guesswork.

Therefore, in this work, we study collisions of 1\,mm-sized grains at different temperatures of up to 1200\,K approaching high-temperature sticking from the lower-temperature end.
We note, though, that sticking is a necessary condition for compound formation but is not sufficient; just sticking grains together does not automatically make them a compound. Grains also have to fuse together on the relevant cooling timescales of minutes to hours. 
Therefore, we support the collision experiments by static heating experiments of the same grains brought in contact with each other at temperatures of up to 1500\,K in order to evaluate the relationship between sticking and compound formation. 

\begin{figure}[h]
	\includegraphics[width=\columnwidth]{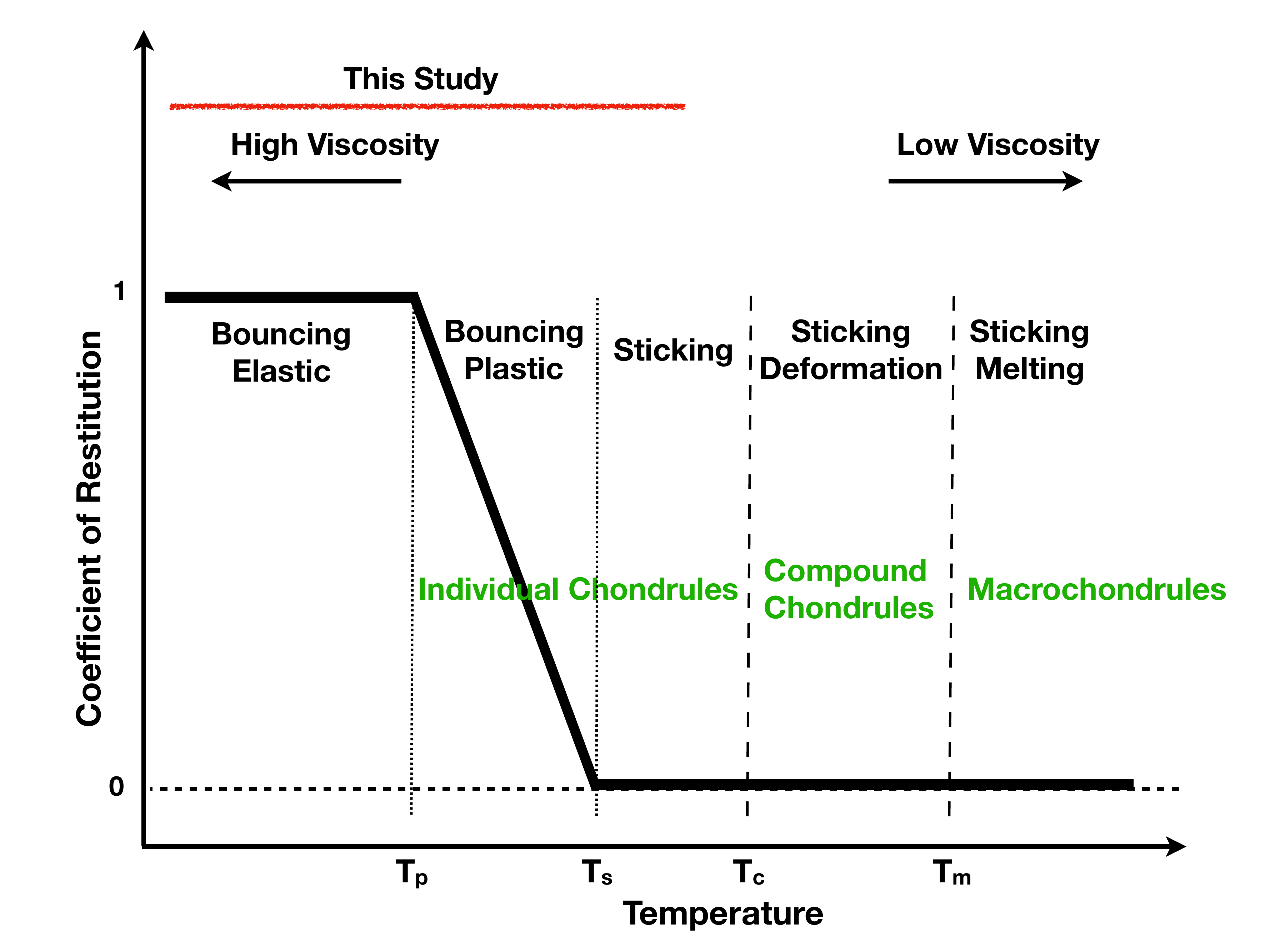}
	\caption{\label{fig.concept}Expected outcomes if chondrules collide at one given velocity.}
\end{figure}

Our expectations of collisional outcomes are shown schematically in Fig. \ref{fig.concept}.
We define a coefficient of restitution $k$, which represents the ratio between the rebound velocity and the collision velocity.
In addition to the temperature, other factors such as the impact parameter and the size of the colliding particles have a large influence on the coefficient of restitution and sticking; e.g., the energy in the center of mass system depends on them. In addition, sticking at grazing incidence becomes less likely. These parameters were not varied in this first study, as the setup was designed for grains to vertically impact a flat target.
If $k=0$, sticking occurs, which is a necessary requirement for compound chondrule formation. If $k\neq0$, no compound can form even if the collision is plastic, i.e., dissipating energy. 
However, chondrules would become slower after a collision. Usually, the gas-grain friction time competes with the collisional time scale. The friction time determines how long it takes a grain to follow the gas motion, i.e., slow down in a gas at rest or adapt to a present gas motion. The friction time can be estimated as $\tau_f = \rho_p / \rho_g \cdot d /c$ with the particle and gas densities $\rho$, the grain diameter $d$ and the sound speed $c$. The collision time is the typcial time that occurs between two collisions. If the grains couple with the gas quickly, the remaining collision velocity is determined by their motion within the gas, e.g., radial drift or turbulence. If they collide before coupling with the gas, the velocities are determined by the collisions and the energy loss in a collision. 
In that case, chondrule velocities might also be determined by bouncing and the detailed coefficient of restitution rather than gas drag. Rough estimates for 1\,mm-sized grains of density $\rho_p = \rm 10^3\,kg/m^3$ at $\rho_g = \rm 10^{-6}\,kg/m^3$ gas density (minimum mass solar nebula at about 1 AU \citep{Hayashi1985}) at $c = 1000\,\rm m/s$ have friction times on the order of 1000\,s. This would be on the same order of the cooling time, and thus for compound chondrules on the same order of the collision timescale. However, small changes in gas density can easily shift the balance to one side or the other. In addition, some care should be taken with gas-grain coupling in general. If compounds form in very dense regions in the sense that the dust-to-gas-mass ratio is much larger than 1, then a chondrule is no longer a "test" particle and the chondrule motion determines the gas motion and coupling via gas is much slower.

We roughly mark different temperatures where changes in collisions occur from bouncing elastic to plastic ($T_p$) to sticking ($T_s$) to compound formation ($T_c$). 
If compounds fused together completely, the result might be a single, larger chondrule, so frequent collisions might lead to the formation of macrochondrules \citep{Weyrauch2012,Bischoff2017}. We mark this as $T_m$.  A specific curve in this temperature diagram would only be valid for one given collision velocity. The relative threshold values will vary with collision velocity, i.e., the sticking threshold will be shifted. If collisions are slower, grains will already stick at lower temperatures. If collisions are too fast, even droplets will not stick together and $T_s$ will be higher than $T_c$. The transition temperatures also vary with cooling time, i.e., if a particle has a longer time span available (hours instead of minutes) to fuse with another particle it has stuck to, then $T_c$ might be somewhat lower. 
We should emphasize, though, that there are more aspects to the characterization of compounds, e.g., with respect to textures, which we do not discuss here. This would add more detail to the problem than could reasonably be shown in a simple diagram; we are only interested here in the limits of compound formation to estimate number densities.

\section{Experiments}

The basic experiments presented here are collision experiments at high temperatures. The collision experiments are conducted with particles resting in a horizontal ceramic tube for typically 10~minutes during their heating before they are launched by tilting the tube. Gravity then accelerates the grains over a distance of some centimeters, resulting in the typical collision velocity on the order of 1\,m/s. In the present study, we start by observing collisions of the hot grains with a flat and cold glass target.
The maximum temperature accessible in the experiments naturally occurs if launching becomes impossible as particles become plastic and stick to the substrate that they are launched from. If grains do not fall by themselves, they are pushed by a feedthrough to reach the highest possible temperature with the given setup.

\subsection{Setup}

\begin{figure}[h]
	\includegraphics[width=\columnwidth]{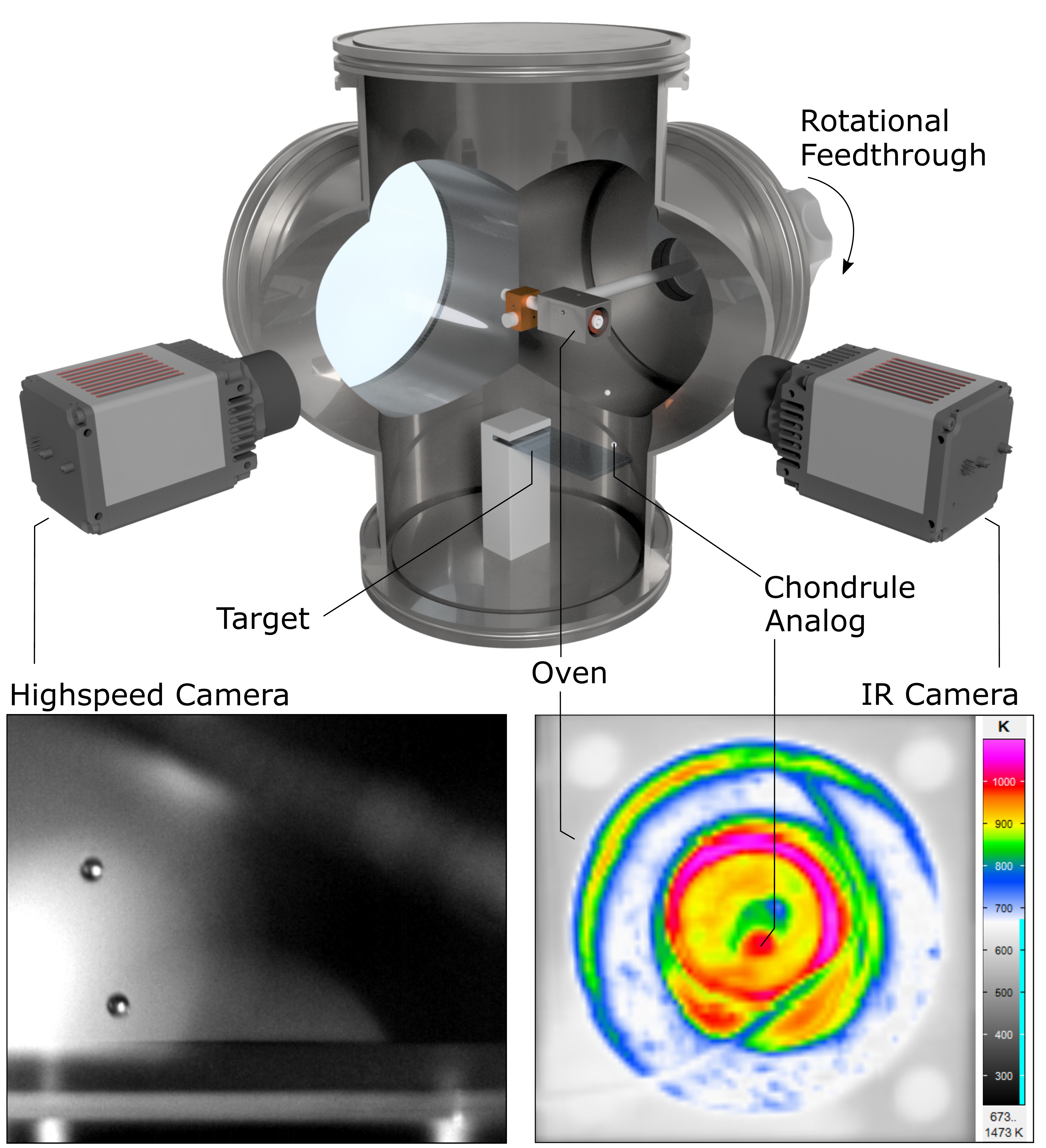}
	\caption{\label{fig.setup}Schematic of the experiment.}
\end{figure}

A schematic of the experiment is shown in Fig.~\ref{fig.setup}.

\begin{figure}[h]
	\includegraphics[width=\columnwidth]{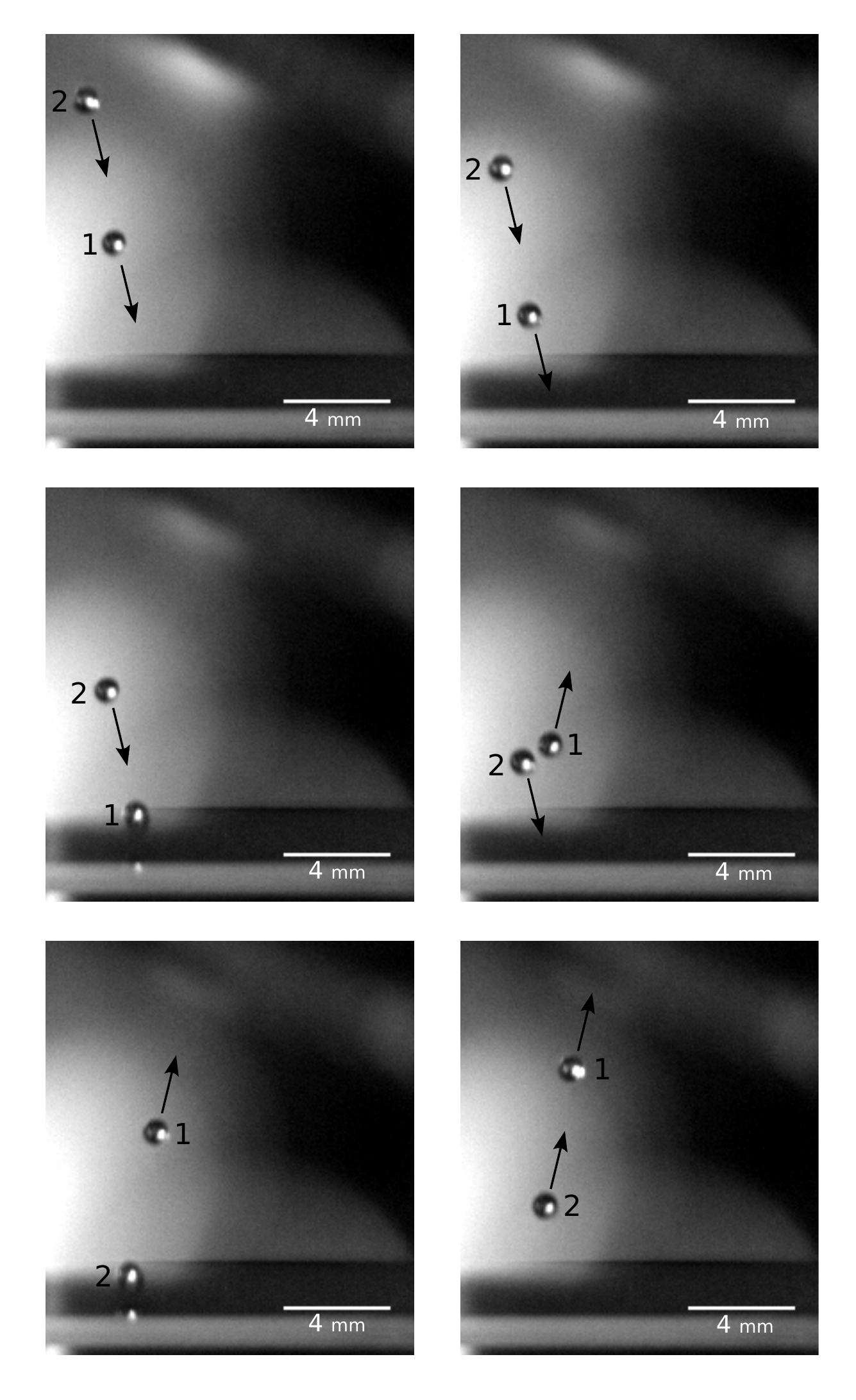}
	\caption{\label{fig.sequence}Two hot glass grains at 973\,K collide with a cold glass surface at room temperature.}
\end{figure}

Particles are observed with two cameras.
The first camera records the falling and rebounding of the grain; an example sequence is
shown in Fig. \ref{fig.sequence}.
The second camera is an infrared camera that measures the temperature upon launch, as indicated in the right-hand inset in Fig. \ref{fig.setup}.

The infrared camera, which was observing through a ZnSe window, was calibrated to the grain temperature by observing grains in an oven at well-defined temperatures.
The experiments are carried out under vacuum (below 1\,Pa). Therefore, cooling during free fall can only proceed through thermal radiation. This cooling is only on the order of a few K, though, and thus negligible in this experiment.

\section{Coefficient of Restitution}

\begin{figure}[h]
	\includegraphics[width=\columnwidth]{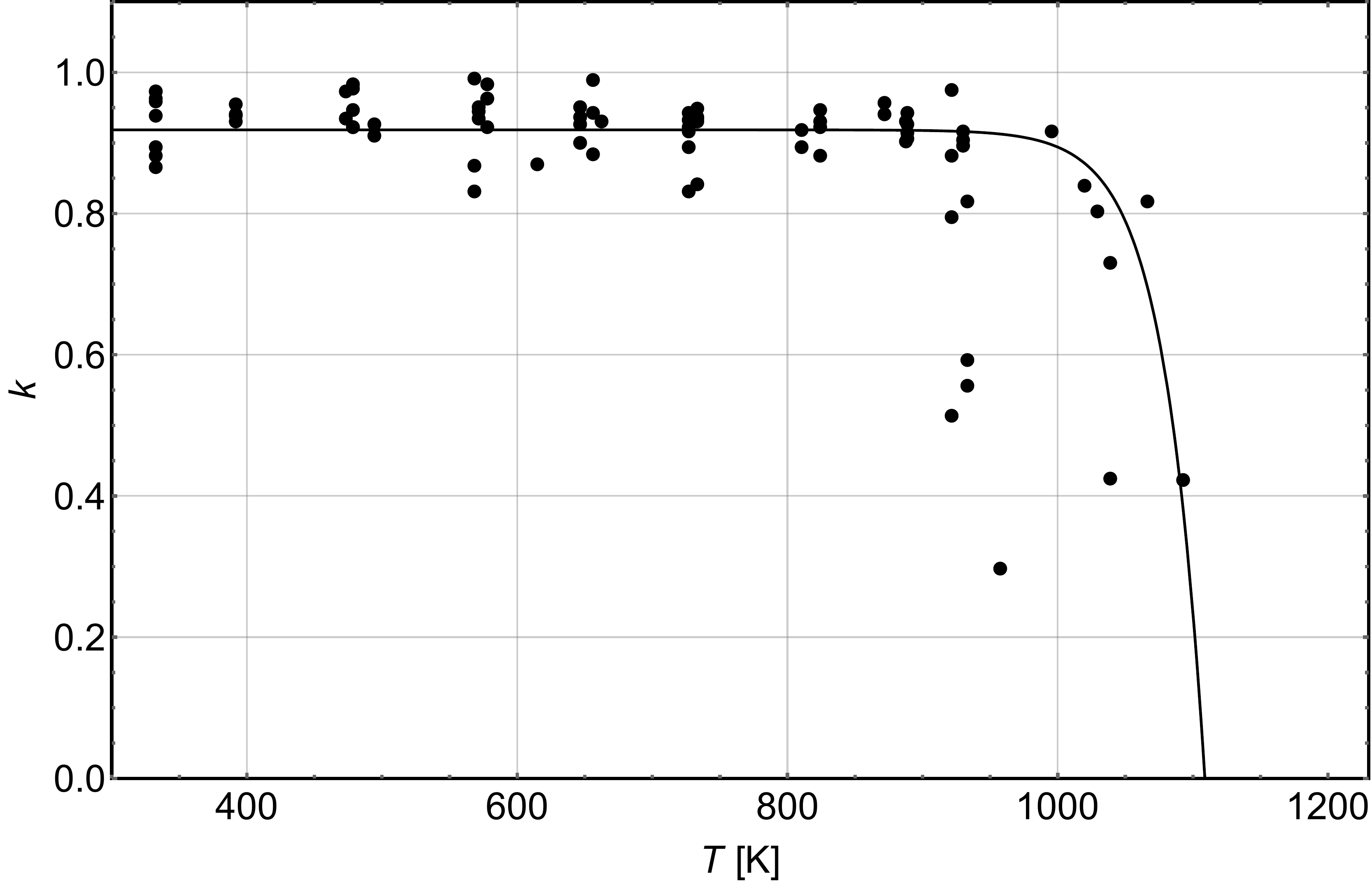}
	\caption{\label{fig.data}Coefficients of restitution for 1\,mm-diameter glass spheres at 1\,m/s impact. The few low values below 1000\,K are likely due to prolonged heating and deformation changing the shape of the grains.}
\end{figure}

\begin{figure}[h]
	\includegraphics[width=\columnwidth]{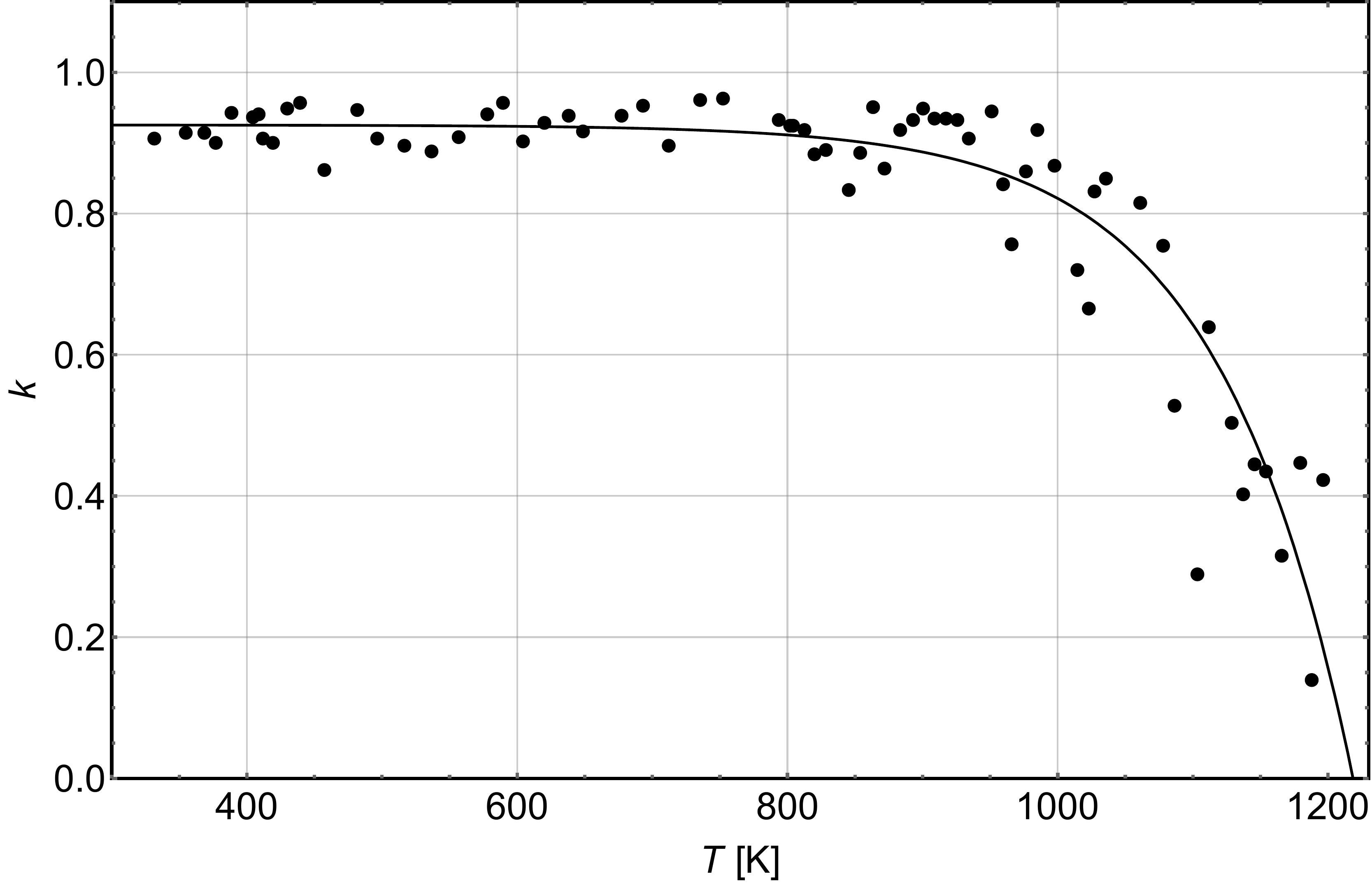}
	\caption{\label{fig.data2}Coefficients of restitution for 1\,mm basalt spheres at 0.8\,m/s impact.}
\end{figure}

We used 1\,mm spherical glass and basalt particles as sample. The basalt spheres are composed of 34\,\% SiO$_{2}$, 14\,\% Al$_{2}$O$_{3}$, 14\,\% Fe$_{2}$O$_{3}$, 13\,\% CaO, 8.5\,\% MgO, 3.5\,\% Na$_{2}$O/K$_{2}$ and 4\,\% other compounds (Whitehouse Scientific Ltd Technical Data Sheet). We did not measure grain sizes within the basalt, as we did not consider this detail important for this study.
Thus, results for chondrules might vary somewhat, but chondrules also have a large variety of liquidus temperatures beyond 1500\,K \citep{Radomsky1990, Zanda2004}. Thus, we especially consider the basalt grains as one possible analog to show the systematics of sticking collisions at high temperatures for chondrules.
Figures \ref{fig.data} and \ref{fig.data2} show the coefficients of restitution, $k$, for all collisions depending on temperature. An exponential fit, $k=a-b \, \exp{(T-c)} $ with $a$, $b$ and $c$ being constants, is included to guide the eye without further underlying physical interpretation at the moment. 

As can be seen, cold particles are very elastic, with high coefficients of restitution. Only at about 900\,K is there a transition where grains start to lose energy in a collision. The further decrease in $k$ occurs over roughly 200\,K up to 1100\,K for glass and over 300\,K up to 1200\,K for basalt. At these temperatures, the heating softens the particles sufficiently so that they stick to the ceramic heating tube and do not fall easily.
The experiment is therefore limiting itself to below the hot and viscous state where sticking might occur after a collision. From the dependence of $k$, we can scale to a first lower limit $T_s$ of temperatures below which compound chondrules cannot form at velocities above 1\,m/s. We note that we did not observe sticking collisions. Therefore, the sticking velocity might be somewhat lower for the deduced $T_s$ or the sticking temperature higher for the given collision velocity. Nevertheless, the data provides significant constraints on sticking.

For slower impacts of collisions of hotter particles, levitation mechanisms are needed to avoid contact with a launcher which is up to future work.

\section{High-Temperature Compound Formation}

\begin{figure}[h]
	\includegraphics[width=\columnwidth]{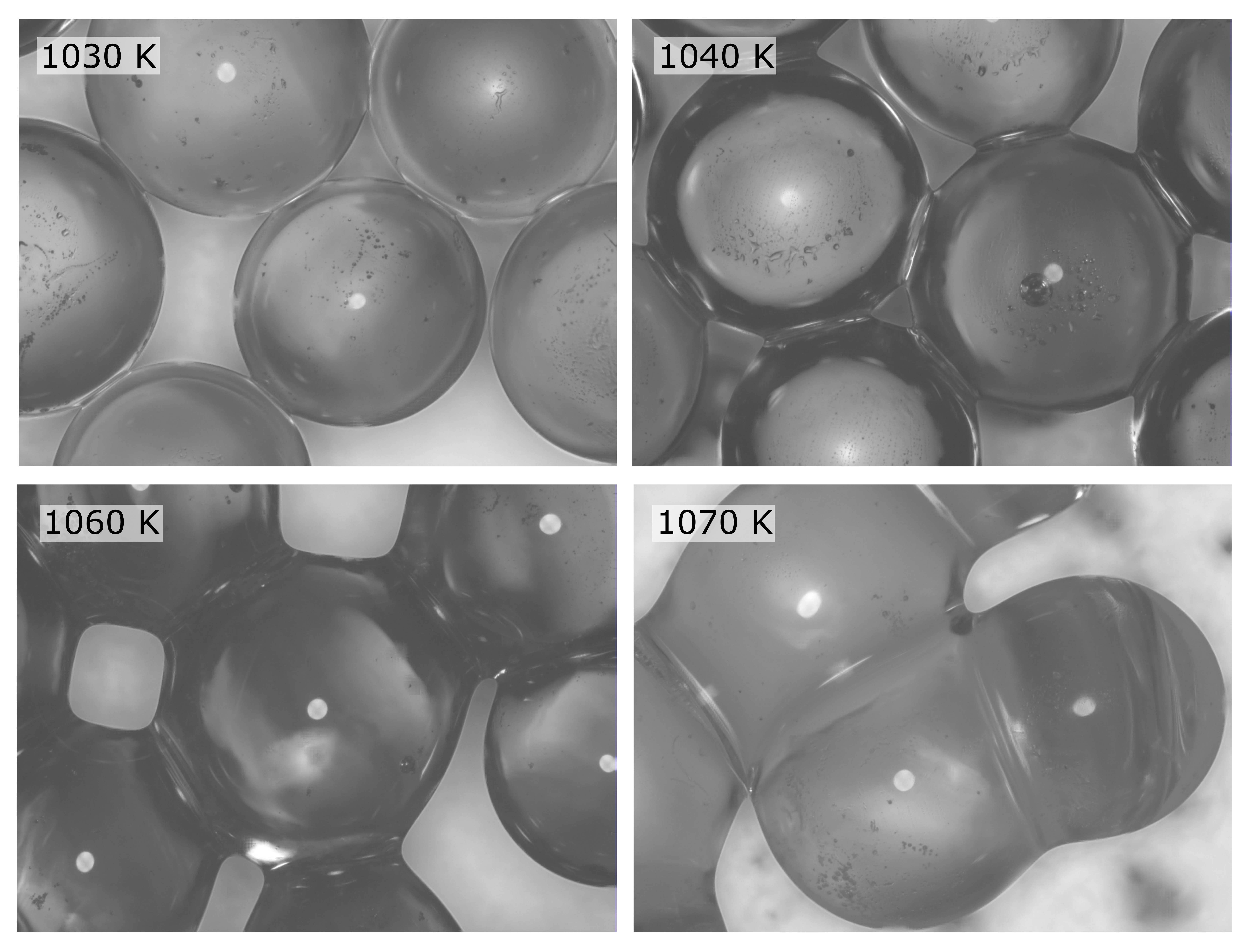}
	\caption{\label{fig.melt1}Images of glass spheres with a mean diameter of 1\,mm heated in an oven at different temperatures for 15\,min.}
\end{figure}

\begin{figure}[h]
	\includegraphics[width=\columnwidth]{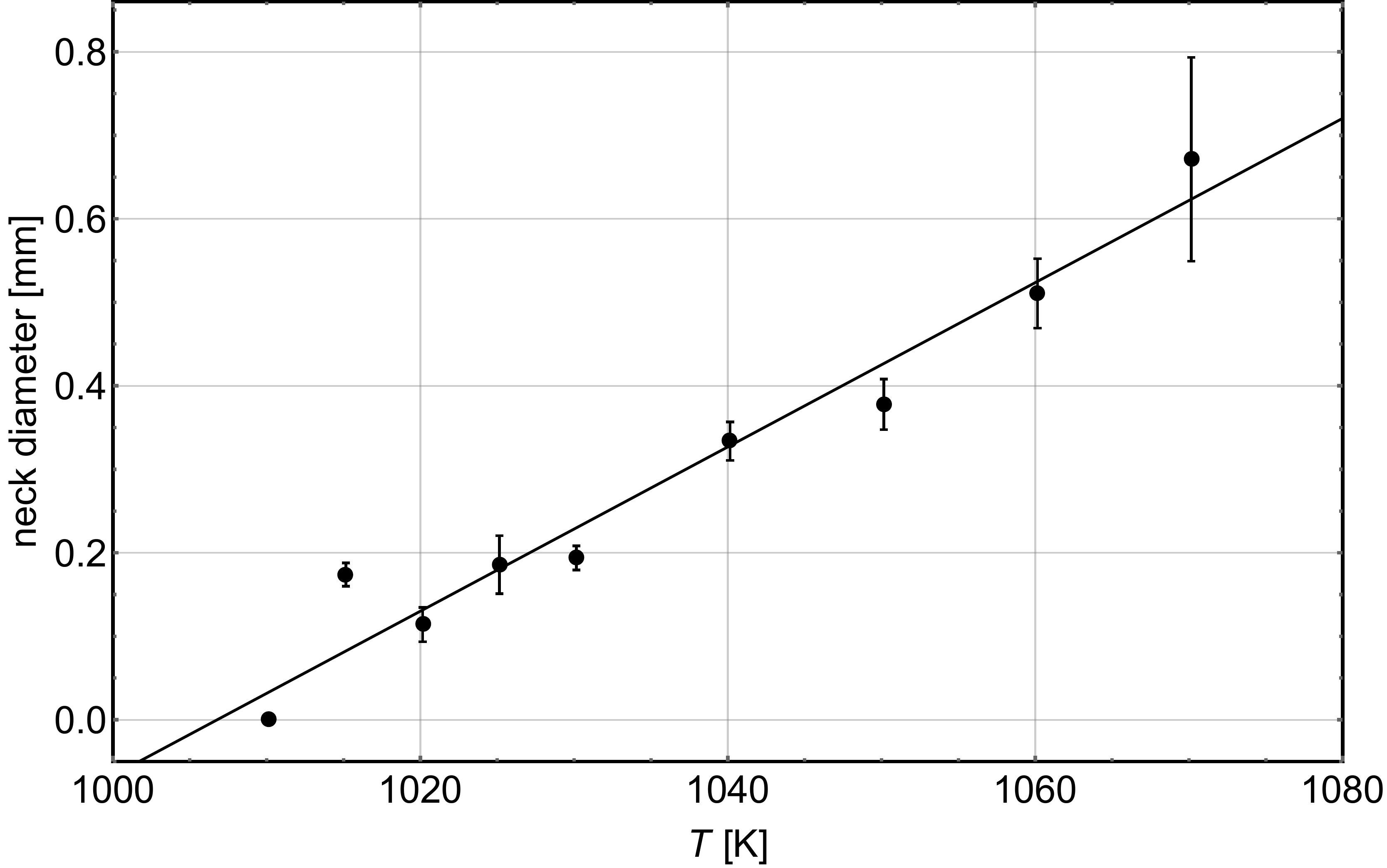}
	\caption{\label{fig.melt1a}Size of the contact area of two glass spheres heated for 15\,min at different temperatures. Each data point represents the mean value of more than 20 contact areas, including the standard deviation.}
\end{figure}

To go slightly beyond the sticking threshold, we additionally performed heating experiments with the same glass and basalt particles used for the collision experiments. The samples were placed in a preheated, external oven at air for different timespans at temperatures beyond their sticking temperature.
Since the sintering process depends on the pressure applied, well-defined initial conditions are needed. In the solar nebula, only cohesion would be present initially, and no external pressure would be present. We therefore also prepared the spheres as a loose ensemble on a flat surface so that they just touched each other without any additional pressure.
The glass is very sensitive to temperature variations. In one series we heated the glass spheres for only 15\,min between 1000 and 1100\,K. It is clearly visible in Fig. \ref{fig.melt1} that glass particles easily sinter together and form necks in the given temperature and time range. At higher temperatures, they melt together 
leaving nothing but a trace of the original grain surface between them. 
This can be quantified in terms of the formed neck diameter, as shown in Fig. \ref{fig.melt1a}.

Even if we do not consider the glass to be representative of chondrules, this shows the scheme of possible transitions. Here, sticking is only expected at about 1100\,K. Grains would form compounds already at 1040\,K if we consider 0.3\,mm as the criterion (see below). By definition, $T_c$ cannot be lower than $T_s$. However, if collision speeds were smaller, $T_c$ might shift to as low as 1000\,K or even lower. 

\begin{figure}[h]
	\includegraphics[width=\columnwidth]{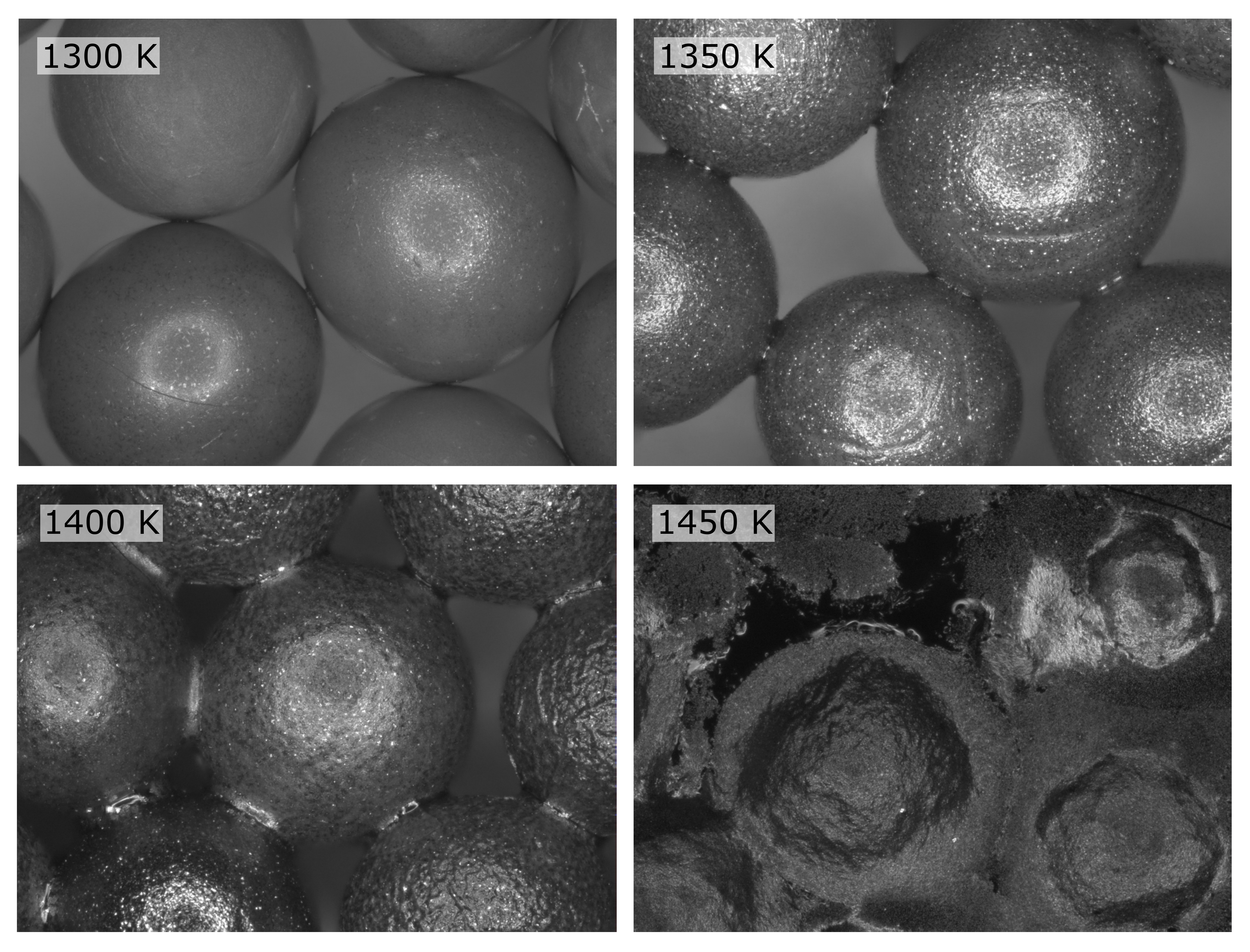}
	\caption{\label{fig.melt2}Images of basalt spheres with a mean diameter of 1\,mm heated in an oven for 24\,h at different temperatures.}
\end{figure}

By contrast, basalt is less sensitive to temperature variations.
Figure \ref{fig.melt2} shows the basalt spheres as they evolved in an oven heated for 24\,h at varying temperatures. 

\begin{figure}[h]
\includegraphics[width=\columnwidth]{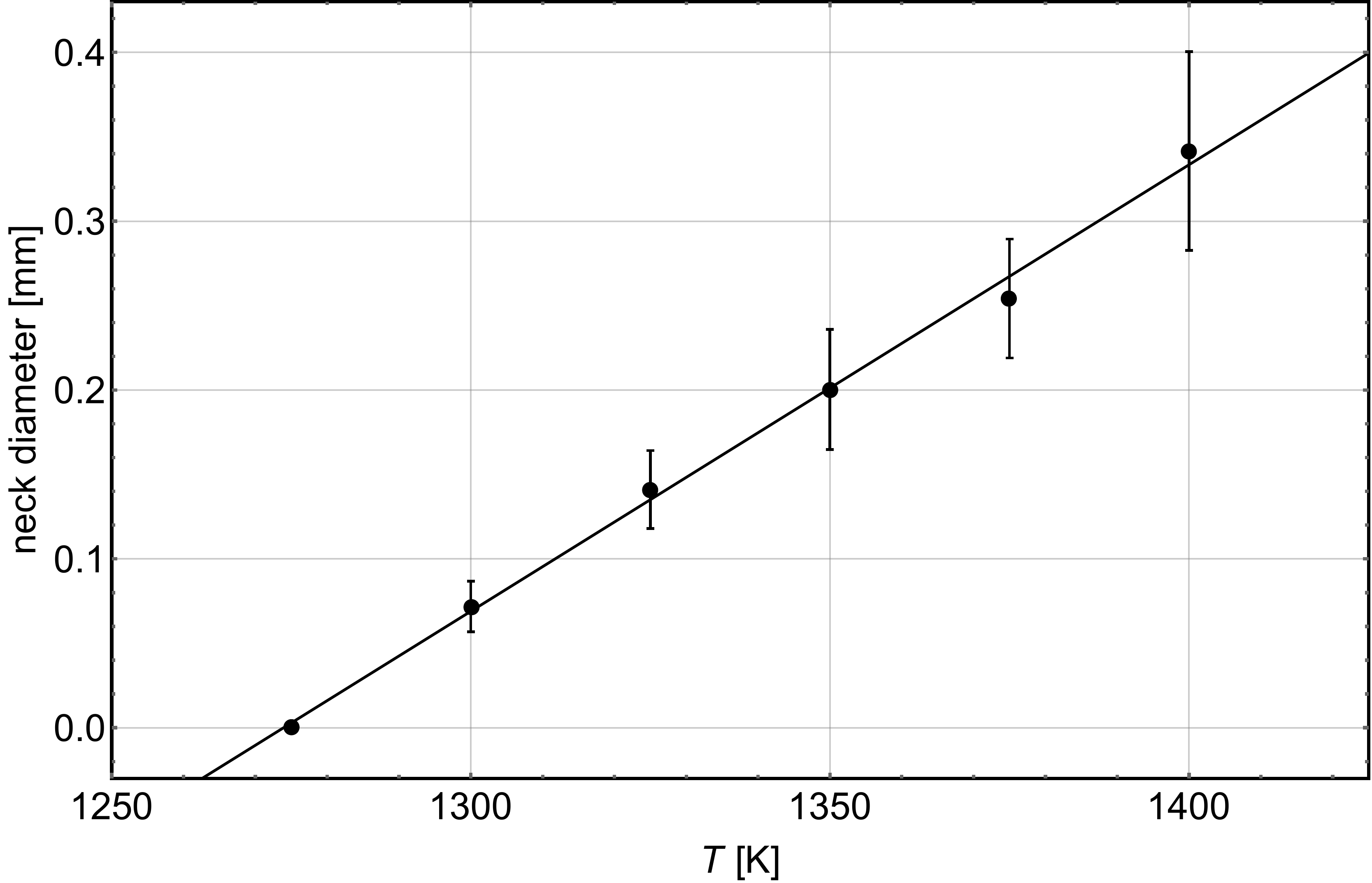}
	\caption{\label{fig.melt2b}Size of the contact area of two basalt spheres heated for 24\,h at different temperatures. Each data point represents the mean value of more than 10 contact areas, including the standard deviation.}
\end{figure}

If the cooling process of compounds is restricted to less than 1~day, then
the lower limit for basalt compounds would be about 1400\,K. This depends, however, on the definition one would choose for a compound chondrule and thus might be somewhat arbitrary. Particles heated below 1400\,K can already have necks up to 1/3 of the particle size after 1~day. However, only at 1400\,K, there is a steep transition where chondrules can have necks on the order of their size before they lose their identity completely at 1500\,K. Thus, we use 0.3\,mm as the approximate neck size for 24\,h heating as the definition for a compound here.
Although we note, that basalt changes its composition as it is heated and that the detailed composition will have an influence on the sticking properties \citep{Demirci2017}, we do not consider this here.
In addition, while the dependence of the neck size on temperature follows a linear trend below complete melting as seen in Fig. \ref{fig.melt2b}, we only consider a fixed heating duration here, although certainly, the neck size is dependent on the duration of heating $\Delta t_{heat}$. \citet{Sirono2011} deduced a time dependence of the neck size for icy particles as $\propto \Delta t_{heat}^{1/4}$, which is a rather weak dependence. This implies that the difference between cooling for discussed timescales between, e.g., 1\,h or 1~day is only on the order of a factor of 2 in neck diameter.

Above 1200\,K but below 1400\,K, basalt grains will not form compounds even if they stick together.
Above 1500\,K, basalt grains melt and fuse together on timescales of minutes in our experiments. This marks the upper temperature limit for compound formation unless cooling occurs very rapidly.

\section{Discussion}

Based on our experiments, collisions of 1\,mm basalt grains will not lead to sticking at 1\,m/s if the particles are cooled to below a critical temperature \mbox{$T_s = 1200\,\rm K$}. Basalt-like compounds are likely only possible above \mbox{$T_C = 1400\,\rm K$}. Only then a significant neck forms, which might count as a compound chondrule, but that threshold is slightly uncertain in principle, as the length of cooling time available is important. 
The same uncertainty holds for the upper threshold. Beyond \mbox{$T_m = 1500\,\rm K$}, the individual features blur rapidly (in terms of their shape).

This outcome essentially contradicts the cold fusion model assumed for compound formation by \citet{Hubbard2015}. He argued that the coalescence time of chondrules at temperatures as low as 1025\,K could be a few minutes. However, even in the case of the glass spheres, only the onset of fusion might occur that cold. For more realistic compositions, coalescence will not occur on short timescales at such a low temperature.

We slightly varied the fall height for the basalt grains, but even at velocities as low as 0.7\,m/s, no sticking collisions occur in the range of the transition temperatures for basalt. Therefore, even if the coefficient of restitution suggests sticking at a certain temperature, the minimum velocity for sticking could not be determined with the given setup. Currently, we only have an upper limit of 0.7\,m/s. In the case of uncharged particles, cold grains of the given size would stick at mm/s, according to \citet{Jungmann}. 

Regarding the Weber number as the ratio between impact energy and surface energy for completely molten droplets, \citet{Ciesla2006} argues that the critical velocity before a droplet is destroyed might be about 1.4\,m/s. Sticking might be possible up to these velocities as mentioned in \citet{Hubbard2015}. The glass grains that show the capability of fusing together right at the sticking threshold indicate that sticking occurs at velocities only below  1\,m/s. For still higher temperatures (with a lower surface tension), the sticking velocities would increase and might approach the critical velocity. However, such compounds would only exist briefly, as they would just fuse together. In any case, there is obviously not a continously monotonous trend of increasing sticking velocity with increasing temperature as destruction occurs, eventually. As \citet{Ciesla2006} argues, a liquid shell and more solid core could lead to an increase in sticking velocity. This would be interesting to study in future collision experiments.

The model for compound chondrule formation underlying our experiments is a standard one. Chondrules are assumed to cool after their formation and collide during cooling \citep{gooding1981}.
If these collisions lead to sticking, and if time allows the formation of necks, fusing them together, we call this compound formation. 
We note that there are other mechanisms discussed for compound chondrule formation. \citet{Hubbard2015} suggests them to be fused rather cold, although certainly not all compounds where textures of compounds are related to each other can be explained that way, and this would take a long time as outlined above. \citet{Sanders1994} consider an eruption model where a secondary chondrule emerges from the crack of a primary chondrule. In addition, enveloping compounds exist, which might be explained by a relic chondrule in a flash molten porous aggregate \citep{Wasson1995}. Collisions, eruptions and enveloping mechanisms for different compounds are also considered in a study by \citet{Akaki2005}.

To test the limits, we carried out a large number of collisions between spherical 1\,mm glass and 1\,mm basalt grains and a flat glass target at different temperatures. 
The experiments show that at somewhat above 900\,K both kinds of grains start to dissipate energy in collisions due to viscous deformation, and the collisions get less elastic. 
This is important in simulating a cloud of chondrules colliding while still warm. The collisional cooling depends on the coefficient of restitution.

The glass grains make a rapid transition to complete melting, leaving a rather small temperature range and time window for glass compounds to form. While basalt is still not very ductile beyond the sticking threshold we deduced, the glass already sinters rapidly. The glass grain experiments suggest sticking velocities to be below 1\,m/s in general.

For basalt, we found that the coefficient of restitution only drops to 0 at 1200\,K for velocities smaller than 0.7\,m/s, although no sticking could actually be observed.  Significant visible fusion of grains occurs only at 1400\,K on timescales of 1 day. Only above that temperature do real compounds --- based on neck formation --- eventually form after a collision. The upper temperature separating compound formation from complete melting and possibly macrochondrule formation is 1500\,K. Here, particles in contact fuse together completely in 1\,h, making distinctions between the individual particles impossible. 

We would like to note as a disclaimer that the experiments should certainly be repeated with real chondrules. With real chondrules, the liquidus temperatures might not only be different but also strongly depend on the composition of the material \citep{Radomsky1990, Zanda2004}. Changes also have to be expected if the target is a second hot grain instead of a cold flat target, and size variations in grains might also change the picture somewhat. Nevertheless, the experiments suggest that 1\,mm compound chondrules only form in a small temperature window of 100\,K. In addition, collisions have to be below 1\,m/s (with a larger target) to lead to sticking in the first place as a requirement for compound formation. 
These numbers include some unknowns and collision experiments at higher temperatures are needed, but this requires different setups for free grains.

If compound chondrules are a measure of local density within the solar nebula,
then these results can be used to pinpoint densities in the disk. 
An additional point to be made is that, details aside, not all hot chondrules colliding will lead to the formation of compound chondrules as, e.g., the sticking probabilities are just  not always averaged over all possible collisions that occur with varying temperature, impact angle, compositon or chondrule sizes. This has essentially already been emphasized in earlier works, e.g., by \citet{gooding1981} or \citet{Ciesla2006}. If not all collisions lead to compounds, this implies that only a lower-number density of chondrules can be deduced from the number of compounds found.
For better results, all details have to be known, from the cooling curve to the yet unknown sticking velocity dependence on temperature. 
We consider our work only as a first step in studying high-temperature collisions to obtain some better constraints on sticking properties. However, if the temperature range is only small (100\,K) and sticking velocities are only low, the re-interpretation of compound chondrule statistics might already result in higher chondrule number densities than previously estimated. For example, \citet{Bischoff2017} assumed a sticking velocity of 5\,m/s, which might have been too conservative. The already-extreme density might therefore even be higher by a factor of a few.

This still includes a large factor of uncertainty, but the experiments pave the way for  providing better estimates on particle densities in the solar nebula during early phases of planet formation. 

One might also ask what the effect of chondrule collisions will be that collide at higher temperatures outside of the compound temperature window. Again, this will depend on the collision velocity. If this starts off high, the droplets will likely be destroyed. Such considerations, e.g., lead to the sticking velocity limit of about 1 m/s introduced by \citet{Kring1991} or \citet{Ciesla2004}. Then, initial aggregates and initial chondrules would have been somewhat larger. However, if the droplets collide more gently, they would form one larger chondrule in a collision that might evolve to macrochondrules after several collisions. If two chondrules for some reason would have different temperatures, certainly either enveloped compounds or the destruction of the liquid chondrule are possible. The experiments reported here cannot yet answer in what direction this will go. However, future collision experiments at higher temperatures will allow some of the speculations to disappear.

\section*{Acknowledgements}
M.K. is supported by the DFG under grant WU~321/14-1. This work is also embedded in the project funded by DFG grant WU~321/18-1. We would also like to thank the two anonymous reviewers.

\section*{References}

\bibliography{bib}

\end{document}